# Single-molecule interfacial electron transfer dynamics manipulated by external electric current

**Guofeng Zhang, Liantuan Xiao*, Ruiyun Chen, Yan Gao, Xiaobo Wang and Suotang Jia**

Interfacial electron transfer (IET) dynamics in 1,1'-dioctadecyl-3, 3, 3', 3'-tetramethylindodicarbocyanine (DiD) dye molecules / indium tin oxide (ITO) film system have been probed at the ensemble and single-molecule level by recording the change of fluorescence emission intensity. By comparing the difference of the external electric current (EEC) dependence of lifetime and intensity for enambles and single molecules, it is shown that the single-molecule probe can effciely demonstrate the IET dynamics. The backward electron transfer and electron transfer of ground state induce the single molecules fluorescence quenching when an EEC is applied to ITO film.

## Ⅰ. Introduction

Interfacial electron transfer (IET) dynamics play an important role in many chemical and biological processes.[1-5] However, IET processes are usually very complex due to high dependence on its local environments. Single-molecule spectroscopy has led to many surprises and has now become a standard technique to study complex structures[6] or dynamics[7-10] including of photoinduced, excited-state intramolecular, and IET dynamics. The optical signals of single molecules provide information about dynamics of their nanoscale environment, free from space and time averaging. Single-molecule studies of photoinduced electron transfers in the enzyme flavin reductase have revealed formational fluctuation at multiple time scales.[2] Single-molecule studies of photosensitized electron transfers on dye containing nanoparticles also showed fluorescence fluctuations and blinking, and the fluctuation dynamics were found to be inhomogeneous from molecule to molecule and from time to time.[11] Recently developed single-molecule spectroelectrochemistry extends single-molecule approaches to ground state IET by simultaneously modulating the electrochemical potential while detecting single molecule fluorescence.[5,12,13]

Indium tin oxide (ITO) films are the most widely used material as a transparent electrode of organic light emitting diode and also in other devices like solar cells.[14-18] It is interesting to study electron transfer of ITO to suitably modify interactions at the interfaces of dissimilar materials so that desired electronic properties of devices incorporating them can be realized. Single dye molecules dispersed on the semiconductor surface of ITO were used to measure IET from excited cresyl violet molecules to the conduction band of ITO or energetically accessible surface electronic states under ambient conditions by using a far-field fluorescence microscope, and single-molecule exhibited a single-exponential electron transfer kinetics.[19] Here we apply an external electric current (EEC) to manipulate the IET rate between single dye molecules and its neighboring ITO nanoparticles by probing the fluorescence intensity change of individually immobilized single dye molecules dispersed in ITO film.

## Ⅱ. Experimental Section

Cover glass substrates were cleaned by acetone, soap solution, milliQ water sonication and irradiation with ultraviolet lamp. The samples were prepared by spin coating (3000 rpm) a solution of 1,1'-dioctadecyl-3, 3, 3', 3'-tetramethylindodi-carbocyanine (DiD, $10^{-9}$ M, Molecular Probes) in chlorobenzene onto the cover glass substrate. The chemical structure of DiD molecule is shown in Fig. 1(a). The Indium tin oxide (ITO) was purchased from Sigma-Aldrich (Product Number: 700460, dispersion, <100 nm particle size (DLS), 30 wt. % in isopropanol, composition: $In_2O_3$ 90%, $SnO_2$ 10%). ITO film in hundreds of nanometer thicknesses was spin-coated onto dye molecules. Two aluminum leads were fixed to the ITO film, and the interval between the two leads is about 4 mm. After vacuum-dried, the samples were further covered with a poly-(methyl methacrylate) (PMMA) (Mw=15,000, Tg=82 °C, Aldrich) film in order to insulate oxygen. The samples were subsequently annealed in vacuum at 350 K for 5 h to remove residual solvent, oxygen and to relax influences of the spin coating technique on the polymer conformations. The method is depicted schematically in Fig. 1(b).

A 70 picosecond pulse diode laser of λ=635 nm (PicoQuant, PDL808) with a repetition rate of 40 MHz was used to excite single dye molecules. The output of the pulse laser was passed through a polarizing beamsplitter cubes (New Focus 5811) to obtain linear polarization light. A 1/4 wave-plate was used to change the polarized laser into circular polarization light. The laser beam was sent into a conventional inverted fluorescence microscope (Nikon ECLIPSE TE2000-U) from its back side, reflected by a dichroic mirror (BrightLine, Semrock, Di01-R635-25x36), and focused by an oil immersion objective lens

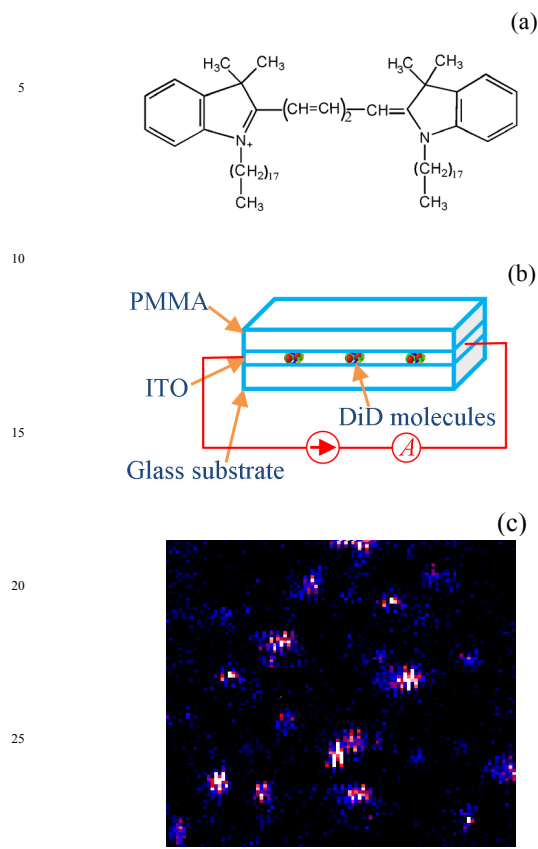

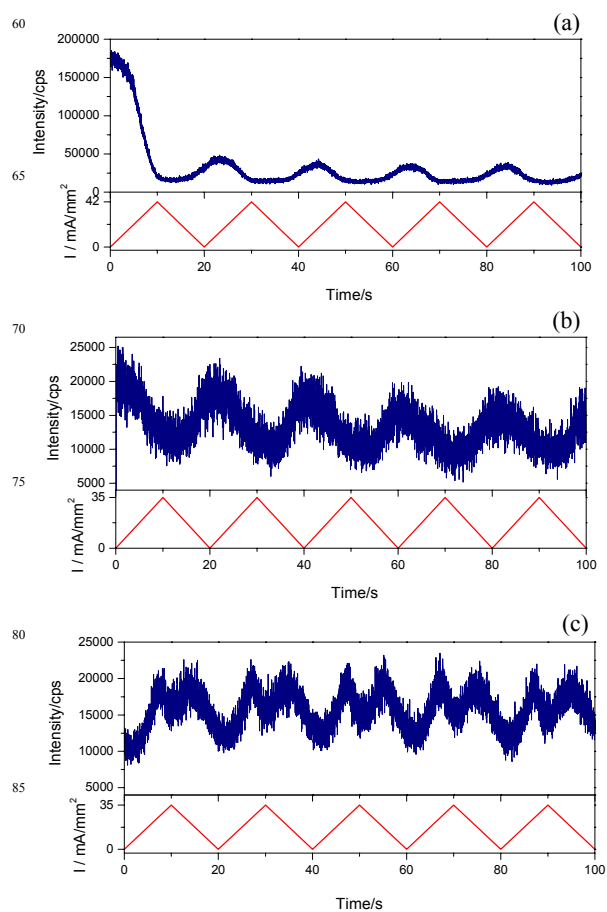

**Fig. 1** (a) Structure of the 1, 1'-dioctadecyl-3, 3, 3', 3'-tetramethylindodicarbocyanine (DiD) dyes molecule. (b) Schematic structure of the sample preparation. (c) Confocally scanned fluorescence image (10 μm×10 μm) of DiD molecules dispersed in an indium tin oxide (ITO) film. Each bright feature may be attributed to a single DiD molecule.

(Nikon, 100×, 1.3 NA) onto the upper sample surface of the cover glass substrate. The fluorescence was collected by the same objective lens and then passed through the dichroic mirror, an emission filter (BrightLine, Semrock, FF01-642/LP-25-D), and a notch filter (BrightLine, Semrock, NF03-633E-25), is focused onto a 100 μm pinhole for spatial filtering to reject out-of-focus photons. Fluorescence photons were subsequently focused through a lens and collected by a single-photon detector (PerkinElmer, SPCM-AQR-15). A piezo-scan stage (Piezosystem jena, Tritor 200/20 SG) with an active x-y-z feedback loop mounted on the inversion microscope was used to scan the sample over the focus of the excitation spot, producing a two-dimensional fluorescence imaging. All measurements were conducted in a dark compartment at room temperature. We use an alternative power source to supply electric current into the ITO film. The applied electric current is proportional to the amplitude of applied external bias voltage, with the relation being $I = 0.87U$. The dye molecules embedded in the ITO film are distinct from the molecules in electric field experiments.[20] In the electric field experiments, a large electric field intensity of about 100V/μm is needed, and dyes must be insulated from the electrode. In our experiment, dye molecules directly contact with ITO nanoparticles for IET controlling with much low voltage needed (less than 0.02 V/μm). In order to distinguish with the normal electric field experiments, we discuss the results ensue from EECs here.

Fig. 2 Three typical patterns of ensemble fluorescence intensity trajectories that were obtained while repeatedly applying a triangle wave EEC sequence to ITO shown by the bottom red curves. (a) The fluorescence intensity shows a fast quenching while applying the EEC to the ITO. (b) The fluorescence intensity shows a decrease while applying the EEC to the ITO. (c) The trajectory shows that fluorescence increases at smaller EEC and then decreases at larger EEC.

## Ⅲ. Results and discussion

### A Fluorescence imaging of single dye molecules in ITO film

Fluorescence imaging using a confocal arrangement has superior sensitivity for spectroscopic measurements and is most suitable for studying single-molecule behavior in dilute sample. Fig. 1(c) displays the confocal fluorescence image of the single DiD molecules within a 10 μm×10 μm area, which is obtained by scanning a sample containing randomly placed isolated single fluorescent molecules dispersed in an ITO film. The imaging is taken in 225 s (150 pixels × 150 pixels), with a pixel integration time of 10 ms. The single molecules are excited with laser excitation intensity of 1.5 kW/cm$^2$ at 40 MHz. The concentration of the DiD molecules was kept at such a low level that either one molecule or no molecules were in the focus. The bright features in the image then represent the fluorescence from individual molecules. The full width at half maximum of typical spots is about 0.3 μm. Meanwhile, it is also found that the spots have different intensities. The intensity variation between the molecules is due to different molecular orientations and environments.[21,22] We have found that the average intensities of the molecules were about an order of magnitude weaker than that of DiD molecules in PMMA without ITO under the same experimental conditions. The weaker signal intensities result from the molecules undergoing the IET with surrounding ITO nanoparticles.[19]

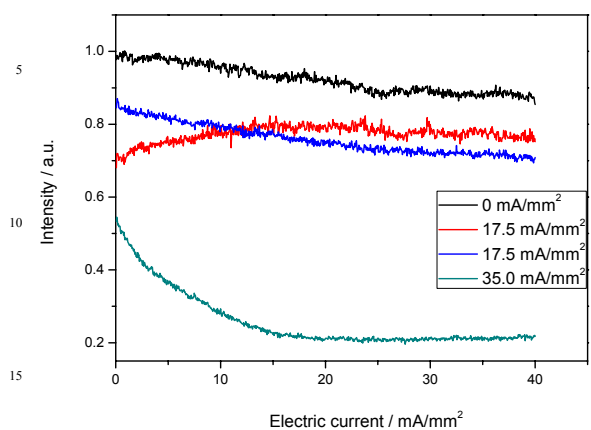

Fig. 3 Normalized fluorescence intensity trajectories of 25 ensembles obtained at different EECs. The black curve shows the average of 25 ensemble trajectories obtained at zero current; the blue curve shows the average of 16 ensemble trajectories with fluorescence quenching obtained at the current of 17.5 mA/mm$^2$; the red curve shows the average of 9 ensemble trajectories with fluorescence enhancement obtained at the current of 17.5 mA/mm$^2$; the green curve shows the average of 25 ensemble trajectories obtained at the current of 35.0 mA/mm$^2$.

**B Electric current response of the ensemble fluorescence**

To study the IET dynamics between dye molecules and ITO nanoparticles, we placed dyes embedded in ITO film. An EEC was applied to the ITO film to manipulate the IET rate. The molecules fluorescence intensity was measured by centering ensemble or one molecule in the laser focus and recording the transient fluorescence intensity, while a time-dependent EEC was applied to ITO film.

We measured the EEC response of the ensemble at different dyes concentration, and the ensemble averaged study would establish how the average fluorescence intensity and lifetime are affected by the electrical current. The Fig. 2 shows three typical ensemble average results. Fig. 2(a) shows that the fluorescence intensity of an ensemble decreases rapidly while applying the current to ITO film. Fig. 2(b) shows a fluorescence quenching trajectory of an ensemble under the EEC. however, fluorescence can be enhanced sometimes for some ensembles at relative small EECs as shown in Fig. 2(c). The similar fluorescence enhancement was also observed in the Ref. 23, which was explained that potential-induced modulation of the excited state reduction processes (i.e., electron transfer from ITO to the polymer) dominates the low-potential fluorescence-modulation effect. We can find from the Fig. 2 that the ensembles do not show complete fluorescence quenching at the relative large external electric current. This may be due to that those dye molecules in the ensemble are not in good contact with the ITO nanoparticles, which shows poor IET under these conditions. Fig. 3 shows normalized fluorescence intensity trajectories of 25 ensembles with different dyes concentration obtained under different EECs. The black curve shows the average of 25 ensemble trajectories obtained at zero current. The red curve and blue curve are constructed by sorting trajectories with fluorescence enhancement and fluorescence quenching at the current of 17.5 mA/mm$^2$. In our experiment, for approximately 30% of the ensemble data show fluorescence enhancement effects at relative small current. The green curve is the average of 25 ensemble trajectories obtained at the current of 35.0 mA/mm$^2$, and it shows fluorescence quenching of the ensemble at a large current. The various responses to EECs arise from the heterogeneity of site-specific molecules.

**C Electric current response of single-molecule fluorescence**

We have detected several hundreds of single DiD molecules and all of the molecules sensitively responded to EEC. Fig. 4 shows the typical fluorescence emission for different individual DiD molecules in dependence of the EEC applied to the ITO film as a function of time. Fig. 4(a) is the fluorescence intensity time trace of one DiD molecule as an EEC of 32.0 mA/mm$^2$ periodically applied to the ITO film, which shows that the EEC can quench effectively the fluorescence emission of single-molecule. Fluorescence blinking observed in the trace shows a single DiD molecule emission and the blinking may be related to the triplet state and/or charge transfer between single molecule and its local environment.[24] While the electric current of 32.0 mA/mm$^2$ is applied to the electrodes, the fluorescence intensities of the single molecule exhibit an exponential decay with the time constant about 2.24 ± 0.23 s. The fluorescence emission gradually recovers to the initial value which needs about 10 s after switching off the EEC. Both of the decay time and the recovering time depend on the molecular neighboring environment. The emission intensities of another single molecule at different electric currents are showed in Fig. 4(b), note that the rates of intensity decay are different, when electric currents are applied between 10 s and 40 s. These intensity decay traces were fitted by single-exponential function with the time constants of 5.80 ± 1.20 s, 3.20 ± 0.54 s and 1.38 ± 0.09 s under the electric current of 17.3 mA/mm$^2$, 39.0 mA/mm$^2$, and 46.8 mA/cm$^2$ respectively. Note that the bigger current, the faster response time. The intensities recover to the initial values which need about 7.5 s after switching off the EEC. It is also found from the Fig. 4(b) that the 46.8 mA/mm$^2$ current (and even bigger) can quench almost completely the fluorescence emission. The two molecules have the different recovering times, which arises from the heterogeneity of site-specific molecular interactions.

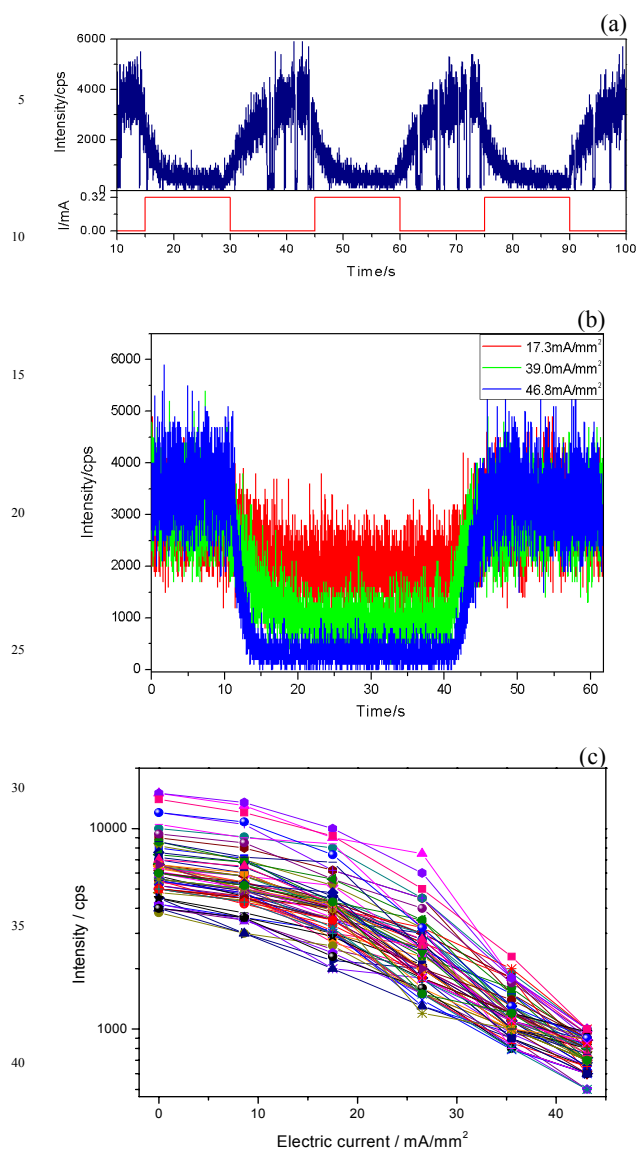

**Fig. 4** Electric current response of DiD single-molecule emission intensity. (a) One single molecule fluorescence emission patterns as electric current applied periodically to the electrodes. (b) Another single-molecule emission intensities under different electric currents applied between 10 s and 40 s. (c) Single-molecule fluorescence intensity trajectories (data points) for 75 DiD single molecules that were obtained while applying different currents.

In addition, we record the average fluorescence intensities of each single DiD molecule at different currents. The statistical data of 75 DiD single molecules are shown in Fig. 4(c). All the DiD single molecules exhibit intensity decrease with thelarge EEC.

We present a model, as shown in the Fig.5, to explain the results of fluorescence quenching of single-molecule. The schematic representation of energy levels, basic photoinduced and electric current driving processes, and multiple electron transfer cycles in DiD / ITO system are shown.

When an EEC is applied to the ITO film, the Fermi level of the ITO is tuned by the potential. With a positive potential, the Fermi level of ITO is decreased and there is more driving force for the forward electric transfer (FET) and the electron transfer of ground state (GET), but backward electron transfer (BET) is suppressed simultaneously. While the BET is suppressed completely at a large EEC, the fluorescence is quenched completely. When EEC is switched off, the Fermi level of ITO will gradually recover to its original value and the driving force for the FET and GET will decrease, simultaneously the BET will increase, thus the fluorescence will recover gradually.

### D  Electric current dependence of fluorescence lifetime

Lu and Xie have presented that each single molecule exhibits a single-exponential IET dynamics in dye molecules/ITO film system, the rate of FET, $k_{FET}$, can be measured by the fluorescence decay of excited dye molecules. And the changes of fluorescence lifetimes were attributed to the FET.[19, 25-27]

Fig. 6(a) shows three typical fluorescence decay curves for an ensemble (average intensity is about 800k cps), a subensemble (average intensity is about 40k cps) and a single DiD molecule (average intensity is about 8k cps ) in ITO film. The decay curves are fitted with single-exponential decay with the time constant about 1.41 ns, 1.16 ns and 0.90 ns respectively. Accordingly, the lifetimes of DiD single molecule are shorter than the 2.5~3.0 ns lifetime measured in the polymer.[28] Fig. 6(b) shows that the fluorescence lifetimes of ensemble average for 45 ensembles at three different electric currents, ranging from 1.1 to 1.8 ns. The average fluorescence lifetime is 1.43 ns at 0 mA/mm$^2$, 1.40 ns at 17.5 mA/mm$^2$, and 1.38 ns at 35.0 mA/mm$^2$ respectively. Fig. 6(c) shows fluorescence lifetimes of 65 DiD single molecules at three different electric currents, ranging from ~300 ps to 1.4 ns. The average fluorescence lifetime is 0.73 ns at 0 mA/mm$^2$, 0.67 ns at 17.3 mA/mm$^2$, and 0.59 ns at 39.0 mA/mm$^2$ respectively. Unfortunately, we cannot measure a single-molecule lifetime shorter than 300 ps due to the limited sensitivity for time resolution of the instrumental response. As mentioned above, the FET rate is highly sensitive to the

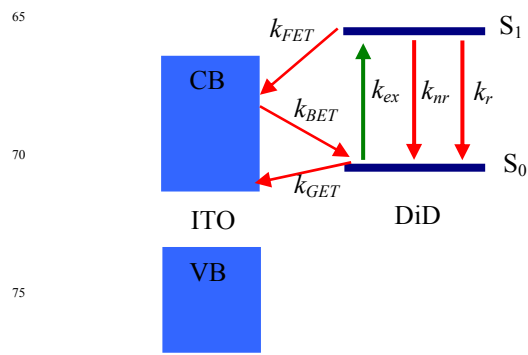

Fig. 5 Schematic representation of energy levels, basic photoinduced and electric current driving process, and multiple electron transfer cycles in DiD / ITO system. $k_r$, radiative decay rate; $k_{nr}$, intrinsic nonradiative decay rate; $k_{FET}$, the rate of forward interfacial electric transfer from excited molecule to semiconductor; $k_{BET}$, the rate of backward electron transfer; $k_{GET}$, the rate of electron transfer from ground state of molecule to ITO; CB, the conduction band of ITO; VB, the valence band of ITO.

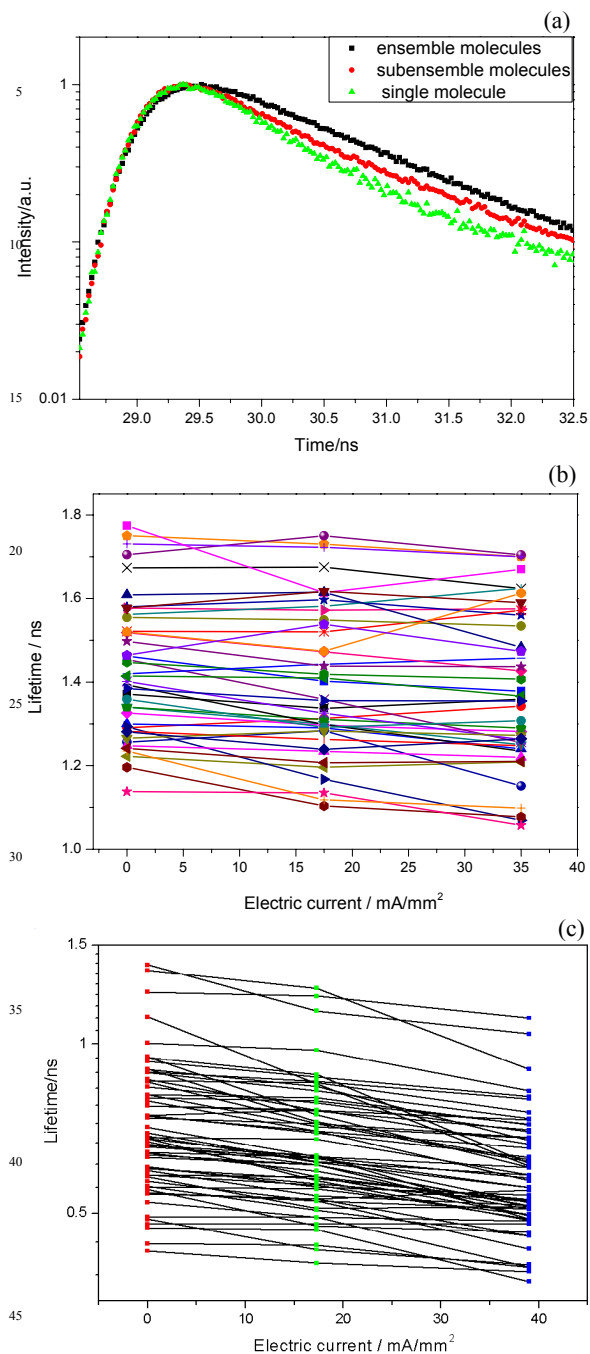

So when some dye molecules in the ensemble that are not in good contact with the ITO nanoparticles which results in poor IET dynamics. Thus, fluorescence lifetime of the ensemble is longer than the lifetime of single-molecule. In the typical dye-photosensitization system, FET can shorten the fluorescence lifetime and reduce fluorescence quantum yield of the dye molecules.[19,25-27] Under the EEC, the small change of fluorescence lifetimes may indicate the change of the FET rate, and the fluorescence quenching may be mainly dominated by the BET and GET.

## IV. Conclusions

Individual DiD dye molecules dispersed in ITO semiconductor films as probes of IET dynamics, which should help to understand the intrinsic properties of electron transfer at interface between organic molecules and transparent semiconductor materials. While the EEC was used to drive the FET and GET, and suppress the BET, the change of $k_{FET}$ induces the change of fluorescence lifetime and the increasing $k_{GET}$ and decreasing $k_{BET}$ would quench the fluorescence. Due to the inhomogenous nature of the interactions from molecule to molecule, the lifetime under EEC is inhomogeneous. The EEC dependence of lifetime distribution clearly demonstrates a manipulated IET dynamics, which can be revealed by the single-molecule experiments better than by the ensemble-averaged measurements. The results could open up a new path to manipulate single-molecule electron transfer dynamics by using EEC while measuring single-molecule fluorescence intensity and lifetime simultaneously.

## Acknowledgments

The project sponsored by the 863 Program (2009AA01Z319), 973 Program (Nos. 2006CB921603 and 2010CB923103), Natural Science Foundation of China (Nos. 10674086 and 10934004), NSFC Project for Excellent Research Team (Grant No. 60821004), TSTIT and TYMIT of Shanxi province, and Shanxi Province Foundation for Returned Scholars. The authors are grateful to the referees for the advice to the paper.

## Notes and references

*State Key Laboratory of Quantum Optics and Quantum Optics Devices, College of Physics and Electronics Engineering, Shanxi University, Taiyuan, 030006, China. Fax: +86 0351 7018927; Tel: +86 0351 7018489; E-mail: xlt@sxu.edu.cn*

**Fig. 6** (a) Fluorescence decays for an ensemble molecules (black curve), a subensemble molecules (red curve) and a single molecule (green curve) in ITO film, measured by time correlated photon counting. The decay curves are fitted with single-exponential decay with the time constant about 1.41 ns, 1.16 ns and 0.90 ns respectively. (b) The fluorescence lifetimes of ensemble average for 45 ensembles at three different electric currents. The average fluorescence lifetime is 1.43 ns at 0 mA/mm$^2$, 1.40 ns at 17.5 mA/mm$^2$, and 1.38 ns at 35.0 mA/mm$^2$ respectively. (c) The fluorescence lifetimes of 65 molecules at three different electric currents. The average fluorescence lifetime is 0.73 ns at 0 mA/mm$^2$, 0.67 ns at 17.3 mA/mm$^2$, and 0.59 ns at 39.0 mA/mm$^2$ respectively.

interactions between the adsorbate dyes and ITO nanoparticles.